\documentclass[10pt, conference, letterpaper]{IEEEtran}
\IEEEoverridecommandlockouts

\usepackage{cite}
\usepackage{booktabs}
\usepackage{amsmath,amssymb,amsfonts}
\usepackage{algorithm}
\usepackage{algpseudocode}
\usepackage{graphicx}
\usepackage{textcomp}
\usepackage{xcolor}
\usepackage{makecell}
\usepackage{marvosym}
\usepackage{url}
\usepackage{hyperref}
\def\BibTeX{{\rm B\kern-.05em{\sc i\kern-.025em b}\kern-.08em
    T\kern-.1667em\lower.7ex\hbox{E}\kern-.125emX}}
\begin{document}

\title{Packet-Level DDoS Data Augmentation Using Dual-Stream Temporal-Field Diffusion}

\author{%
  Gongli Xi\textsuperscript{1,3,4,*},\; Ye Tian\textsuperscript{1,2,*},\; Yannan Hu\textsuperscript{4,\Letter},\; Yuchao Zhang\textsuperscript{2},\; Yapeng Niu\textsuperscript{2},\; Xiangyang Gong\textsuperscript{1,2,\Letter}\\
  \textsuperscript{1}State Key Laboratory of Networking and Switching Technology, Beijing, China\\
  \textsuperscript{2}School of Computer Science (National Pilot Software Engineering School),\\
  Beijing University of Posts and Telecommunications, Beijing, China\\
  \textsuperscript{3}School of Cyberspace Security, Beijing University of Posts and Telecommunications, Beijing, China\\
    \textsuperscript{4}Zhongguancun Laboratory, Beijing, China\\[0.5ex]
 \{kevinxgl, yetian, yczhang, Niuyp, xygong\}@bupt.edu.cn\\
  \textsuperscript{*}Co–first authors. \textsuperscript{\Letter} Corresponding author: huyn@zgclab.edu.cn, xygong@bupt.edu.cn
}

\maketitle

\begin{abstract}
In response to Distributed Denial of Service (DDoS) attacks, recent research efforts increasingly rely on Machine Learning (ML)-based solutions, whose effectiveness largely depends on the quality of labeled training datasets. To address the scarcity of such datasets, data augmentation with synthetic traces is often employed. However, current synthetic trace generation methods struggle to capture the complex temporal patterns and spatial distributions exhibited in emerging DDoS attacks. This results in insufficient resemblance to real traces and unsatisfied detection accuracy when applied to ML tasks.
In this paper, we propose Dual-Stream Temporal-Field Diffusion (DSTF-Diffusion), a multi-view, multi-stream network traffic generative model based on diffusion models, featuring two main streams: The field stream utilizes spatial mapping to bridge network data characteristics with pre-trained realms of stable diffusion models, effectively translating complex network interactions into formats that stable diffusion can process, while the temporal stream adopts a dynamic temporal modeling approach, meticulously capturing the intrinsic temporal patterns of network traffic. Extensive experiments show that our synthetic data more closely matches the original traffic distributions than state-of-the-art methods and significantly boosts detection models’ accuracy in both offline benchmarks and real-world scenarios, while improving their ability to recognize unseen attack patterns. The code is available at \href{https://github.com/Lrbomchz/DSTF_Diffusion}{this link}.

\end{abstract}


\section{Introduction}
\label{sec:intro}
Cybersecurity is one of the most challenging aspects of the information technology field, especially when facing the continuously evolving threat of Distributed Denial of Service (DDoS) attacks. DDoS attacks render network services unavailable by overwhelming target network resources with an influx of requests, resulting in severe consequences.

In response, researchers are employing increasingly sophisticated models to identify DDoS attacks \cite{b33}. For instance, various data-driven approaches \cite{b61, b3, b26, b29, b2} have been extensively utilized in security defense systems. However, the effectiveness of these systems largely depends on the quality and diversity of the data available for training. Although publicly available datasets exist, they are often unrealistic and insufficient to be used as ground truth for characterising DDoS attacks \cite{b47}. Furthermore, the collection and sharing of real network data are frequently constrained by privacy and legal issues \cite{b46}, which further limit the scale and diversity of the datasets.

These challenges can be addressed by data augmentation through generative models that creates new synthetic network traces from existing datasets. By introducing variations and maintaining the inherent characteristics of network traffic at the mean time, this approach enhances both scale and diversity of original datasets \cite{b43 ,b20, b37}. While it is shown that state-of-the-art synthetic traffic generation methods' performance good on common network datasets \cite{b43, b20}, we find them insufficient for producing high-quality DDoS traces.
Unlike common network traces, DDoS attacks not only exhibit \textit{packet-head-field aspect} (e.g. \textit{sip}, \textit{dport}) that identify different attack vectors, but also show \textit{behavior aspect} on temporal dimension and spatial dimension. For example, pulse-wave attacks consist of high-rate short-lived bursts, with each burst leverages a different attack vector \cite{b49}, while carpet bombing targets entire CIDR blocks with multiple attack vectors, rather than individual IP addresses \cite{b48}.

Given the intricacy of DDoS traffic patterns, current generation methods can be divided into two primary paradigms: Generative Adversarial Networks (GANs) based models~\cite{b43}, which engage a generator and discriminator in adversarial training, and diffusion-based models~\cite{b55,b20}, which impose noise on data and learn a corresponding denoising process. GANs often suffer from mode collapse~\cite{b25}, producing low-diversity, repetitive traffic patterns, whereas diffusion methods are hindered by limited attack-specific data and semantic misalignment. Moreover, all existing approaches rely on \textbf{\textit{benign}} traffic for training and consequently fail to capture the temporal dynamics inherent to DDoS attacks.

To overcome the limitations of existing works, we introduce \textbf{D}ual-\textbf{S}tream \textbf{T}emporal-\textbf{F}ield Diffusion (\textbf{DSTF}-Diffusion), a novel multi-view, multi-stream traffic generation framework, which consists of two streams: temporal stream and field stream.
In the field-stream module, DSTF-Diffusion first converts DDoS attack packets into images and automatically generates corresponding packet descriptions (e.g., protocol, attack type). These annotated images are then used to fine-tune the powerful Stable Diffusion foundation model—thereby transferring it to the network domain. However, because the autogenerated descriptions contain numerous network terms that a model pretrained on general real-world imagery cannot interpret, DSTF-Diffusion departs from prior approaches by applying a mapping that replaces each network term with a distinct color name interpretable by diffusion models. In the temporal-generation module, DSTF-Diffusion hierarchically deconstructs the irregular attack time series—first capturing the global attack trajectory, then isolating localized characteristic patterns—and employs a diffusion model to learn both the distribution of individual patterns and the relationships among different patterns. This process uncovers consistent malicious behaviors concealed within otherwise unstructured sequences.

Experimental results affirm that our proposed model excels beyond existing solutions in terms of data fidelity and diversity. Additionally, through meticulous evaluation, we show that our generated datasets significantly improve the practical utility of training data for cybersecurity defense mechanisms, boosting these models’ ability to counter evolving network threats and thereby advancing both cybersecurity research and its real-world deployment.

Our contributions are as follows:
\begin{itemize}
  \item \textbf{Dual-Stream Trace Generation Network:} We propose a unified architecture that jointly models packet‐field semantics and temporal dynamics to synthesize realistic DDoS traffic traces.
  \item \textbf{Field‐Stream Encoding:} Through a semantically aligned mapping framework, we bridge network-specific terminology and real-world semantic concepts, enabling the diffusion model to synthesize DDoS traffic with enhanced fidelity via more fine-grained descriptive cues.
  \item \textbf{Temporal‐Stream Modeling:} We design a diffusion stream to capture sequential patterns and intensity variations in network traffic, enabling faithful simulation of complex attack behaviors.
  \item \textbf{Empirical Validation:} Experiments on benchmarks and real-world settings show marked gains in classification accuracy, feature‐space separation, and generalization to unseen attack types.
\end{itemize}

\section{Background and Motivation}

\subsection{Scarcity of Network Data}
\label{sec:bm:scarcity}
Public datasets in the field of DDoS attack identification are very limited and stale. Commonly used datasets include CAIDA-2007 \cite{b51}, DARPA\cite{b53}, and the CICDDoS-2019\cite{b52}, where CAIDA and DARPA datasets feature real attacks, while the CIC series consists of attacks simulated through artificially constructed topologies. The attack scenarios and types in these dataset have become outdated. Moreover, self-collected datasets often suffer from issues such as small size and data imbalance.

These problems are widespread across various domains of networking. Hence, researching ways to expand datasets to propel network research has become a critical need.

\subsection{Data Augmentation Using Generated Data}
\label{sec:bm:dataaug}
Data augmentation using generated data has proven effective in various fields, especially when there is a lack of labeled data. Efforts have been made to apply these successful experiences to the network domain to expand datasets \cite{b57,b43,b62}. However, these generation methods often focus only on partial information or statistics of the original data, making them unsuitable for complex environments like DDoS attacks. Considering the success of data augmentation using generated data in other networking area \cite{b20}, we believe this approach could also be viable for DDoS attack traffic. Nevertheless, to comprehensively capture the characteristics of DDoS attacks, more precise generation is needed.

\begin{figure}[ht]
    \centerline{\includegraphics[width=0.5\textwidth]{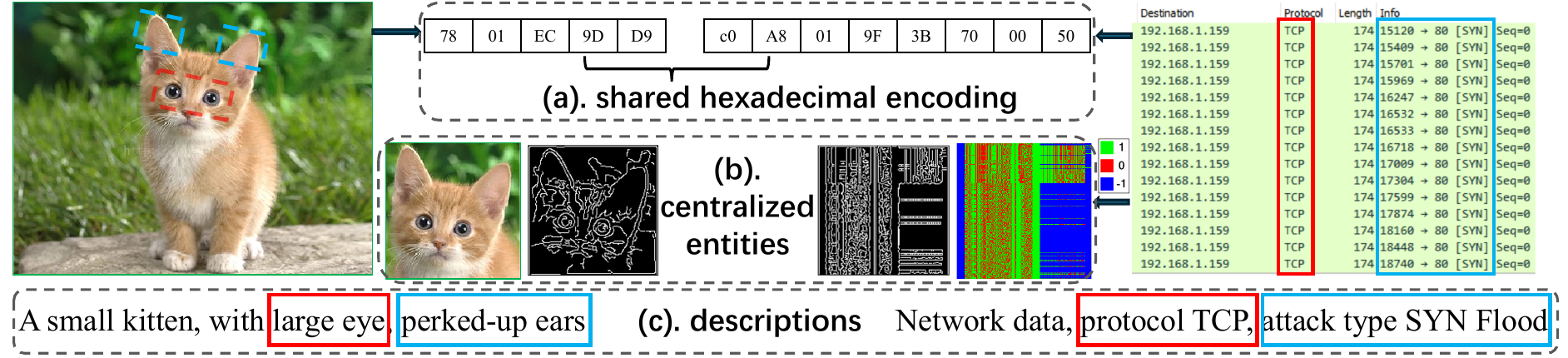}}
    \caption{Motivation for Using Vision Models in DDoS Data Generation}
    \label{fig:moti_field}
\end{figure}
\subsection{Motivation for Using Diffusion Models in DDoS}
\label{sec:bm:challenge}

Diffusion models \cite{b14} intrinsically learn probabilistic data distributions and sample from them. Through this modeling paradigm, they can generate novel instances that differ from the original training examples (diversity) while still satisfying learned constraints (fidelity), thereby enhancing the capabilities of downstream task models.
\begin{figure*}[t]
\centering
\includegraphics[width=0.92\textwidth]{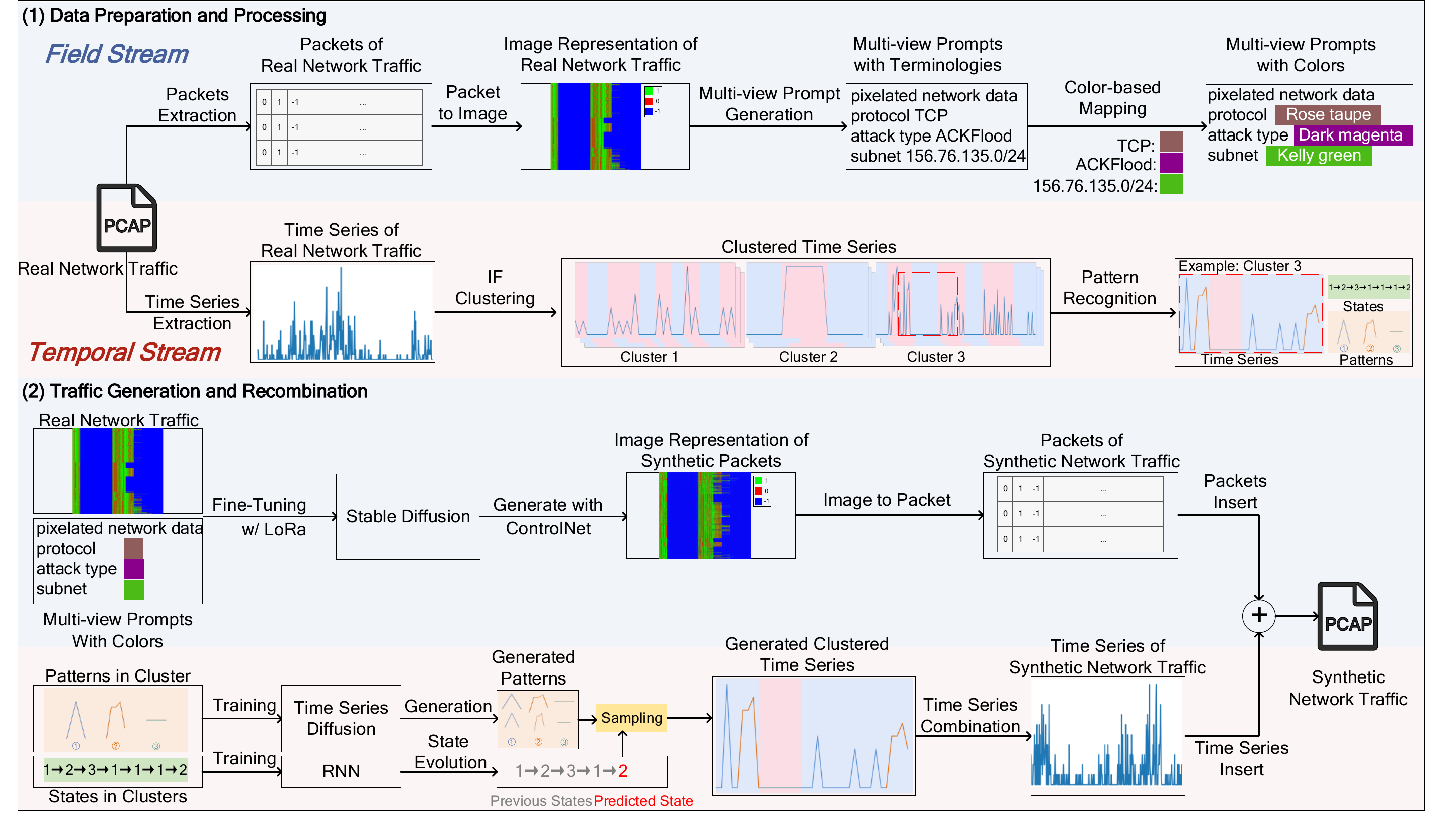} 
\caption{Overview of DSTF-Diffusion. Best viewed in color.}
\label{fig:overview}
\end{figure*}
However, training a diffusion model from scratch to capture the distribution of DDoS attack traffic requires an enormous volume of labeled samples—data that are rarely available in the cybersecurity field.

Fortunately, Stable Diffusion \cite{b34} in the computer-vision domain provides exactly such an ideal initialization.
Several works have transformed network data—such as IPv6 prefixes \cite{b57}, and traffic traces \cite{b20, b37}—into image representations and harnessed Stable Diffusion’s powerful generalization to synthesize realistic artifacts. In the context of DDoS attacks, DSTF-Diffusion’s conversion of packet fields into images is motivated by their shared intrinsic nature.

As illustrated in Fig. \ref{fig:moti_field}: \textbf{(a)} network packets and natural images can both be viewed as hexadecimal data streams; \textbf{(b)} once projected into the image domain, they both exhibit centralized entities; and \textbf{(c)} they possess similar descriptive structures, with packet streams automatically derivable from inherent header information without any manual annotation. These parallelisms reveal a deep structural homology between network traffic and visual data, thereby motivating the application of diffusion-based generative models to synthesize realistic DDoS traffic.

\subsection{Differences from Prior Work}
\label{sec:diff_prior_work}

Recent diffusion-based approaches to network-traffic synthesis can be categorized into time-driven methods that focus exclusively on temporal patterns \cite{b37} and field-driven methods that generate traffic at the service or packet level \cite{b55,b20}.

While both temporal continuity and packet-level granularity are essential for accurate DDoS characterization, existing approaches are limited—time-driven methods \cite{b37} ignore spatial fields, while field-driven methods \cite{b55,b20} overlook temporal dynamics. DSTF-Diffusion addresses this limitation by simultaneously modeling sequential patterns and packet field characteristics.


Furthermore, prior transfer methods overlooked the domain gap between network and real-world data and underexploited Stable Diffusion’s fine-grained generative power. In the field stream, DSTF-Diffusion addresses this limitation by (1) mapping network terms to color and (2) leveraging multi-view prompts.

\section{Preliminaries}
\noindent\textbf{Diffusion models} \cite{b14} gradually corrupt a clean data sample \(\mathbf{x}_0\) into Gaussian noise \(\mathbf{x}_T\) through a fixed forward process:
\begin{equation}
    q(\mathbf{x}_{1:T}\mid \mathbf{x}_0)\;=\;\prod_{t=1}^T \mathcal{N}\bigl(\mathbf{x}_t;\,\sqrt{\alpha_t}\,\mathbf{x}_{t-1},\,(1-\alpha_t)\mathbf{I}\bigr),
\end{equation}
and learn a reverse denoising process
\begin{equation}
    p_\theta(\mathbf{x}_{t-1}\mid \mathbf{x}_t)\;=\;\mathcal{N}\bigl(\mathbf{x}_{t-1};\,\mu_\theta(\mathbf{x}_t,t),\,\Sigma_\theta(\mathbf{x}_t,t)\bigr).
\end{equation}
By optimizing a variational bound on \(\log p_\theta(\mathbf{x}_0)\), these models can recover complex distributions from noise. Their ability to incorporate external signals (e.g.\ packet‐level features) makes them well suited to synthesizing realistic DDoS traffic patterns.

\noindent \textbf{Stable Diffusion} \cite{b34} extends the diffusion framework by operating in a latent embedding space, significantly reducing computational cost. Textual prompts $\mathbf{y}$ are injected via a frozen text encoder, and guide the denoising of latent codes \(\mathbf{z}_t\):
\begin{equation}
    p_\theta(\mathbf{z}_{t-1}\mid \mathbf{z}_t,\mathbf{y})
    =\mathcal{N}\bigl(\mathbf{z}_{t-1};\,\mu_\theta(\mathbf{z}_t,t,\mathrm{Enc}(\mathbf{y})),\,\Sigma_\theta(\mathbf{z}_t,t)\bigr),
\end{equation}
In our setting, we replace text prompts with attack descriptors derived from DDoS traffic, steering generation toward plausible attack scenarios.

\noindent \textbf{Seasonal–Trend decomposition via Loess (STL)} splits a univariate time series $s_t$ into three additive components:
\begin{equation}
    s_t = T_t + S_t + R_t,
\end{equation}
where $T_t$ is the smooth trend (intensity) component, $S_t$ is the periodic (pattern) component, and $R_t$ is the residual noise. STL provides explicit intensity and pattern features that our diffusion model uses to model both the long-term escalation and the localized burst dynamics of complex DDoS traffic—altogether without requiring manual annotation. 

\section{Methodology}

In this section, we first provide an overview of DSTF-Diffusion in Section \ref{sec:math:overview}. Next, we introduce the field stream and temporal stream in detail in Section \ref{sec:math:MvP} and Section \ref{sec:math:host}, respectively.

\subsection{Overview of DSTF-Diffusion}
\label{sec:math:overview}

Figure \ref{fig:overview} shows the overview of DSTF-Diffusion. 
The framework is structured into two streams: field stream for packet-field synthesis via Stable Diffusion fine-tuning, and temporal stream for timestamp/pattern synthesis; a recombination step binds them into full traces.



In the following sections, we will focus on discussing the design of field and temporal stream.

\subsection{DSTF-Diffusion: the field stream}
\label{sec:math:MvP}
In this subsection, we first introduce insights into field stream, followed by a detailed description of its implementation. 
Our insights come from the following two aspects:

\textbf{\textit{Mapping network terms to color:}} Stable Diffusion, rooted in the computer‐vision domain, excels at modeling chromatic information, whereas network‐traffic data is inherently textual or numeric. Consequently, it is imperative to bridge visual modalities and network‐specific semantics. We found that Stable Diffusion exhibits markedly greater sensitivity to color channels than to discrete entities or numbers, reflecting its underlying RGB‐based image‐encoding scheme. Given this observation, the remaining question is how to construct this mapping relationship. Therefore, we propose a mapping structure based on colors to help Diffusion better understand the network. 

\textbf{\textit{Multi-view prompt:}} DDoS data features multi-dimensional characteristics, for example, attacks based on the TCP protocol can include types such as SYN, ACK, etc., based on different TCP flags. Thus, unlike NetDiffusion, which uses a single view prompt to depict network trace data, we prefer to use multiple views to help Diffusion better understand network. The multi view prompts also aligns well with the inherent nature of Stable Diffusion to generate precise images from detailed descriptions.

We describe attacks from three views, denoted as 
$ V = (\text{protocol}, \text{subnet}, \text{attack type}) $. Each view contains different categories, for example, under the protocol view, $C_{V_0}= (\text{TCP}, \text{UDP}, \text{ICMP}) $. 

\noindent \textbf{Mapping network terms to color:}
First, we sort the list of colors based on their RGB values:

\begin{equation}
\text{colors} = \text{sort}(\{({Name}_i, RGB_i)\}).
\end{equation}

For the subnet view, each subnet is mapped to the closest color based on its RGB value. For example, the subnet 153.101.21.0/24 can be mapped as (153, 101, 21) in the RGB space, which corresponds to the color Golden Brown:

\begin{equation}
\text{MP}(C_{subnet}^j) = \min_{(Name, RGB)} \left\{ \text{D}(RGB, C_{subnet}^j) \right\}.
\end{equation}

Here, \(C_{subnet}^j\) represents the j-th subnet.

For other views such as protocol and attack type, we evenly space out and select color names from the sorted list:

\begin{equation}
\label{eq:MvP}
\text{MP}(C_{V_{i}}^{j}) = \text{colors}_{\left\lfloor j \cdot \frac{\text{len}(\text{colors})}{\text{len}(C_{V_{i}})} \right\rfloor}.
\end{equation}

Where \(C_{V_i}^j\)is the j-th category of view \(V_i\) (e.g., specific protocol or attack type), and ${\text{len}(C_{V_{i}})}$ is the number of categories in the view \(V_i\).

\noindent \textbf{Multi-view prompts}: To fully leverage the capability of Stable Diffusion to generate high-quality images through detailed descriptions, we design multi-view prompts, as illustrated in Equation \ref{eq:MVPMT}. It consists of two parts: (1). Global prompt, which is present in all prompts and aims to guide the model to generate images with a network image style. (2). Multi-view prompt, where each view has its own prompt template designed to guide the model's refined generation. An example of a multi-view prompt is shown in Figure \ref{fig:overview}.

\begin{equation}
\text{pmt}_{S} = \text{pmt}_{G} + \sum_{C_{V_i}^j \in S} \text{pmt}_{V_i} \left( MP(C_{V_i}^j) \right).
\label{eq:MVPMT}
\end{equation}

After obtaining the image-prompt pairs, we fine-tune the diffusion model using LoRa \cite{b18}. Details of the fine-tuning process are presented in Section \ref{sec:setup}.
\subsection{DSTF-Diffusion: the temporal stream}
\label{sec:math:host}
In this section, we describe our approach to capture and generate temporal information for DDoS traffic, which is particularly complex in modern DDoS attacks. In real traces, the intensity of attack vectors can fluctuate dramatically. Naively modeling each vector’s raw time series produces synthetic traffic that is overly faithful to the source and lacks the variability required for realistic simulations. An ideal temporal generator must therefore preserve the original distribution while introducing plausible variations.

Consequently, we adopt a \textit{Disassembly, Generation, Recombination} pipeline. As illustrated in Figure \ref{fig:overview}, we split the temporal stream into discrete pattern segments, generate diffusion-based variations for each segment, and then recombine them. The temporal stream operates in three main stages:

\textit{1) Integrated Feature Clustering} (IF Clustering): In this stage, DSTF-Diffusion organizes timestamps from various attacks into time series set $Seqs=\{s_1, s_2, \dots, s_n\}$ and clusters similar time series together to form \textbf{\textit{clustered time-series}}. Subsequently, it creates metadata $(\tau_i, C_i, \text{start}_i, \text{duration}_i)$ for each time series, representing the type of attack, the type of sequence (clustering result), start time, and duration, respectively, and constructs a metadata chain $\mathcal{M}$ based on the start time of each attack sequence.

\textit{2) Time Series Generation}: In this stage, DSTF-Diffusion models each cluster at two levels: pattern generation and temporal modeling. Pattern generation replicates and explores fixed patterns in the data, while temporal modeling constructs the evolution of the patterns over time. These two modeling approaches interact to generate a complete time series, resulting in the \textbf{\textit{generated time series}}.

\textit{3) Time Series Combination}: The purpose of this stage is to generate simulated metadata chain $\hat{\mathcal{M}}$ based on the real attack metadata chain $\mathcal{M}$. DSTF-Diffusion designs three methods for combining generated time series, thereby increasing dataset diversity and meeting combination needs across various scenarios. After combination, DSTF-Diffusion converts these generated time series into \textbf{\textit{timestamps}} and inserts them into the network attack fields generated by the field stream.

Next, we discuss these three stages in detail.
\begin{algorithm}
\caption{Integrated Feature Clustering}
\label{alg:IFC}
\begin{algorithmic}[1]
\Require $Seqs = \{s_1, s_2, \dots, s_n\}$, attack types $\tau$
\Ensure Cluster results $C$, metadata chain $\mathcal{M}$
\State \textbf{Decomposition:}
\For{$i = 1$ to $n$}
    \State $(T_i, S_i, R_i) \gets \text{STL}(s_i)$ \Comment{STL decomposition}
    \State $I_i \gets T_i$ \Comment{Intensity component}
    \State $M_i \gets S_i + R_i$ \Comment{Mode component}
\EndFor
\State \textbf{Clustering:}
\State Compute DTW distance matrix $D_{\text{int}}$ for $\{I_1, I_2, \dots, I_n\}$
\State Compute Euclidean distance matrix $D_{\text{per}}$ using Fourier transforms for $\{M_1, M_2, \dots, M_n\}$
\State $D \gets \sqrt{D_{\text{int}}^2 + D_{\text{per}}^2}$ \Comment{Combined distance matrix}
\State $C = \{c_1, c_2, \dots, c_n\} \gets \text{k-means}(D, k)$
\State \textbf{Metadata Chain Construction:}
\For{$i = 1$ to $n$}
    \State $\text{start}_i \gets min(s_i)$
    \State $\text{duration}_i \gets max(s_i) - min(s_i)$
    \State $m_i \gets (\tau_i, C_i, \text{start}_i, \text{duration}_i)$ \Comment{Metadata for $s_i$}
\EndFor
\State $\mathcal{M} \gets SORT(\{m_1,  \dots, m_n\}, by=\text{start})$ 
\State \Return $C, \mathcal{M}$ \Comment{Return the clusters and metadata chain}

\end{algorithmic}
\end{algorithm}

\noindent \textbf{Integrated Feature Clustering:} In our approach, we aim to cluster sequences with similar attack behaviors into the same category, thus enabling the model to effectively shield influences from other types of behaviors during generation. To achieve this, we propose the Integrated Feature Clustering (\textbf{IF Clustering}) algorithm. This algorithm decomposes the attack sequence into two dimensions: mode component and intensity component, using the features of these two dimensions for joint clustering of the attack sequences in the set $Seqs$. The algorithm is described in Algorithm \ref{alg:IFC}.

\noindent \textbf{Time Series Generation:}
In order to model the continuous characteristics of different attack behaviors in terms of time series, we segment a time series $s_i$ into subsequences $s_{i,m}$ of length \(\alpha_i^m \in [l_{\min}, l_{\max}]\), where \(s_i=[s_{i,1}, \ldots, s_{i,m}, s_{i,M}]\). By using a greedy algorithm and DTW distance, it continuously identifies patterns that occur regularly within the time series, representing each subsequence $s_{i,m}$ as a state triplet \( state_{i,m} = (\boldsymbol{p}_{i,m}, \alpha_{i,m}, \beta_{i,m}) \), and constructs the evolutionary process \( state_{i,m-1} \rightarrow state_{i,m} \). Here, \(\boldsymbol{p}_{i,m}\), \(\alpha_{i,m}\), and \(\beta_{i,m}\) respectively denote the pattern, length, and magnitude of the subsequence $s_{i,m}$.




Next, we train a conditionally guided diffusion network \cite{b19} to generate mode-specific subsequences under different attack patterns. These generated subsequences are then organized and stitched together by an evolution model, forming a complete attack sequence.


\noindent \textbf{Time Series Combination:}
\label{sec:comb}
In the first step of temporal stream, we obtain the metadata chain $\mathcal{M}$ of the original data through IF clustering, which reveals how attackers organize individual attacks into a cohesive whole. To ensure the authenticity and diversity of the data and to meet the needs of different scenarios, we design three methods for generating new metadata chain $\hat{\mathcal{M}}$, where the $i_{th}$ element in $\mathcal{M}$ denotes $m_i=(\hat{\tau}_i,\hat{C}_i,\hat{start}_i,\hat{duration}_i)$.

\textbf{Random Combination}: This combination method is intended for attack prevention. That is, when we do not know how attackers organize attacks, we use a random approach to enhance data diversity. The start time for each metadata is randomly selected from the interval \([0, T - duration_i]\), where \(T\) is the total time that must exceed the duration of any tuple in \(\mathcal{M}\). Thus, each new metadata $\hat{m}_i$ can be expressed as:
\[
\hat{m}_i = (\text{rand}(\tau), \text{rand}(C), \text{rand}(0, T - duration_i), duration_i)
\]

\textbf{Markov-based Combination}: This method is typically suited for scenarios where an attack is underway. We model the attacker's previous behavior to predict their next action. This method assumes correlations between attacks, which is a very common assumption in DDoS attacks \cite{b42, b45}.

For each new element $\hat{m}_i$ in metadata chain, its generation relies on the state of the previous tuple: $\hat{m}_i = \text{Markov}(\hat{m}_{i-1})$

where \(\text{Markov}(\cdot)\) denotes the Markov prediction function that determines the next state based on the previous state.

\textbf{Imitative Combination}: This method allows for strengthening defenses against a specific attacker after an attack has occurred, by mimicking the original organizational method of the attacks, the expression is straightforward: $\hat{\mathcal{M}} = \mathcal{M}$.

In the end, DSTF-Diffusion convert the generated time-series into timestamps according to the generated attack metadata chain \(\hat{\mathcal{M}}\), which includes \(\hat{C}_i\), \(\hat{\text{start}}_i\), and \(\hat{\text{duration}}_i\), and assign them to the generated attack fields designated by \(\hat{\tau}_i\).

\section{Evaluation}
We validate DSTF-Diffusion across multiple datasets to ensure that it can simultaneously guarantee the data fidelity and diversity of the generated data. Initially, we confirm that our generated data exhibits higher statistical similarity to real attack data compared to other packet-level generation methods. Subsequently, we demonstrate that the original data, when augmented with DSTF-Diffusion, achieves enhanced performance across \textbf{offline classification} and \textbf{real-world detection} tasks. In Sec. \ref{sec:real_time}, we demonstrate that DSTF enhances the detection model’s \textbf{generalization to previously unseen} patterns.  
\subsection{Experiment setup}
\label{sec:setup}
\noindent \textbf{Datasets:} In our experiments, we utilize the recent public dataset CIC-DDoS-2019\cite{b52} and our own collected dataset, IMB-DDoS 2023. CIC-DDoS-2019 is a structurally simulated dataset covering seven attack types with a fairly balanced distribution. IMB-DDoS 2023, in contrast, was gathered from real DDoS incidents affecting 50 clients over two days; it comprises nearly 700,000 packets sampled at an 8:1 ratio and includes ten attack types with an inherently imbalanced distribution. While IMB-DDoS 2023 features cutting-edge variants such as carpetnet and pulse-wave, CIC-DDoS-2019 spans a broader range of scenarios (NTP, DNS, LDAP, MSSQL, NetBIOS, SNMP, SSDP, UDP, UDP-Lag, WebDDoS, SYN, TFTP). It is important to note that most real-world DDoS datasets (e.g., CAIDA) are no longer publicly shared.

\noindent \textbf{Synthetic Traffic Generation:} DSTF-Diffusion involves data preprocessing, fine-tuning, and generation with Stable Diffusion. To ensure fairness in comparison, our experimental setup is completely aligned with that of NetDiffusion \cite{b20}. In the data preprocessing stage, we use nPrint \cite{b16} to convert each packet into a 1088-dimensional vector of \{-1, 0, 1\}, and to ensure the independent use of the color channels for each value, we map them to blue (0,0,1), red (1,0,0), and green (0,1,0), respectively. During the fine-tuning phase, we use prompt-image pairs as data and fine-tune the model using LoRa \cite{b18}. In the generation phase, we integrate ControlNet \cite{b44}. ControlNet enables Stable Diffusion to prioritize the generation of images corresponding to specific protocols.

\subsection{Statistical Analysis}
\begin{figure}[ht]
\centering
\includegraphics[width=0.40\textwidth]{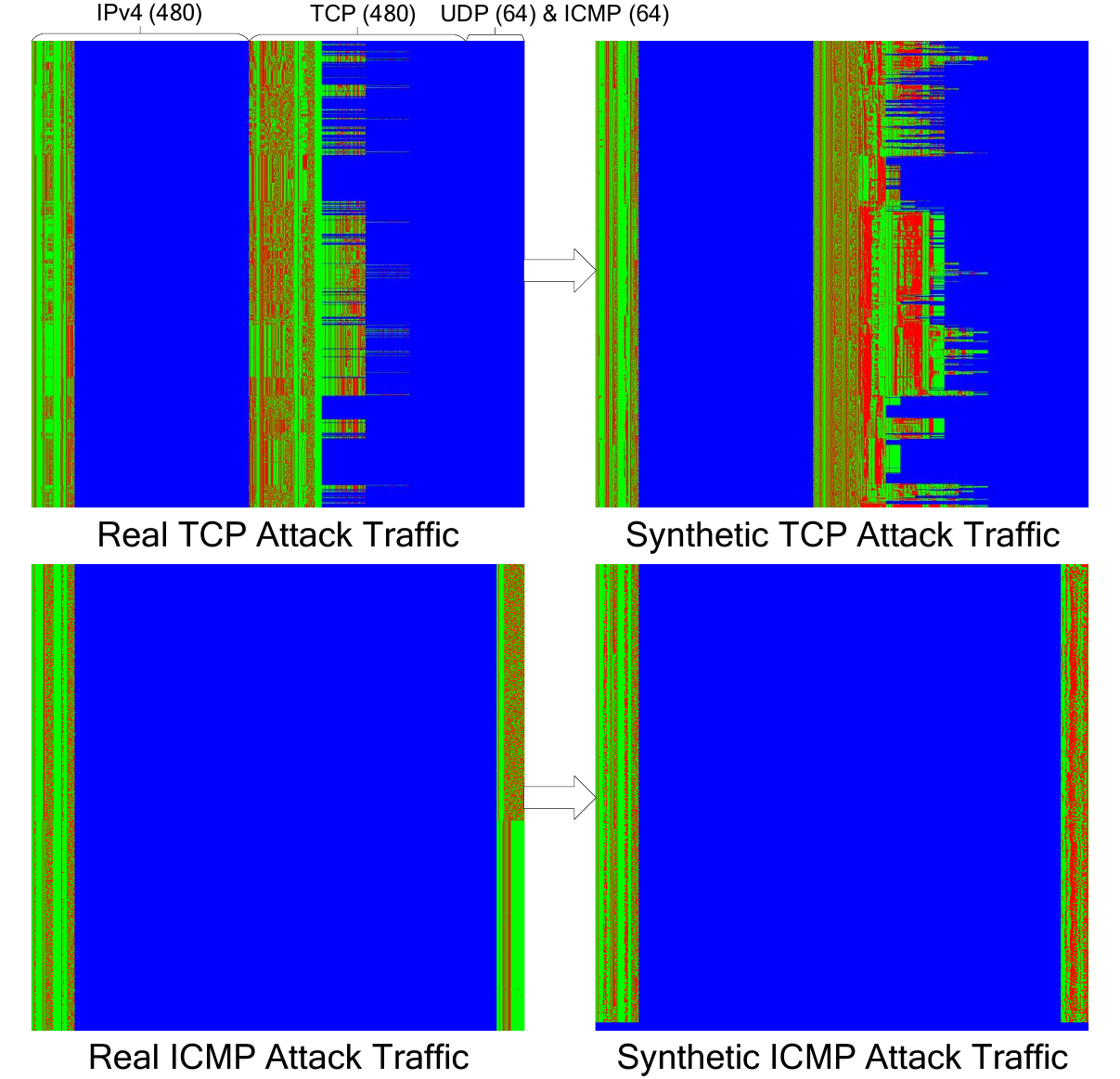} 
\caption{Image Representations of Real and Synthetic Attack Traffic.}
\label{fig:img_rep}
\end{figure}

Figure \ref{fig:img_rep} shows examples of visual representations of generated packets for different protocols. The image size is 1024x1088, where each row represents a packet and each column (or group of columns) corresponds to one network fields transformed by nPrint. The field ranges for different protocols are labeled in the figure.

An important metric for measuring the quality of generated data is its statistical similarity to real data. In our evaluation, we compare with three baselines: (1) \textbf{Random generation}, which represents the worst-case scenario and serves as the lower bound for statistical similarity. (2) \textbf{NetShare} \cite{b43}, which produces synthetic NetFlow attributes and outperforms most of the other GAN-based methods.(3) \textbf{NetDiffusion} \cite{b20}, a generative model based on Stable Diffusion that surpasses other methods in terms of statistical fidelity of the generated data. Specifically, we assess statistical similarity both at a global level (all features) and at a more focused level (protocol).

We employ four different metrics to quantify statistical similarity: Jensen-Shannon Divergence (JSD), Total Variation Distance (TVD), Hellinger Distance (HD) and Validity (Valid). 
Overall, these metrics reveal the statistical similarities between real and synthetic datasets from different perspectives. Note that NetShare generates flow-level features (e.g., duration, packet count), limiting its applicability to analyses requiring detailed packet information. Consequently, it does not provide results for `Protocol' and `Valid' metrics, which are assessed at the packet level.
\begin{table}[ht]
\centering
\caption{Performance Comparison of Traffic Generation Methods Across Different Protocols}
\label{tab:MvP-sta-sim}
\resizebox{.92\columnwidth}{!}{%
\begin{tabular}{l|ccc|ccc|c}
\toprule
 Method & \multicolumn{3}{c}{All Generated Features} & \multicolumn{3}{c}{Protocol} & \multicolumn{1}{c}{Valid} \\
 \midrule
  & JSD & TVD & HD & JSD & TVD & HD &  \\
\midrule
\multicolumn{8}{c}{Protocol \textbf{TCP}} \\
\midrule
 Random & 0.825 & 0.997 & 0.956 & 0.832 & 1.000 & 1.000 & 0 \\
 NetShare & 0.259 & 0.377 & 0.379 & - & - & - & - \\
 NetDif &  0.082 & 0.203 & 0.221 & 0.176 & 0.376 & 0.424 & 0.721 \\
 MV & 0.045 & 0.113 & 0.120 & 0.182 & 0.418 & 0.473 & 0.931 \\
 MP & 0.051 & 0.141 & 0.150 & 0.186 & 0.398 & 0.452 & 0.965 \\
 DSTF & \textbf{0.043} & \textbf{0.111} & \textbf{0.120} &\textbf{ 0.175} & \textbf{0.364} & \textbf{0.417} & \textbf{0.965} \\
\midrule
\multicolumn{8}{c}{Protocol \textbf{UDP}} \\
\midrule
 Random & 0.825 & 0.997 & 0.956 & 0.832 & 1.000 & 1.000 & 0 \\
 NetShare & 0.213 & 0.256 & 0.268 & - & - & - & - \\
 NetDif & 0.522 & 0.830 & 0.822 & 0.367 & 0.668 & 0.698 & 0.443 \\
 MV & 0.081 & 0.172 & 0.185 & 0.115 & 0.274 & 0.281 & 0.723 \\
 MP & 0.207 & 0.451 & 0.471 & 0.260 & 0.497 & 0.541 & 0.982 \\
 DSTF & \textbf{0.046} &\textbf{ 0.101} & \textbf{0.115} & \textbf{0.073} & \textbf{0.195} & \textbf{0.196} & \textbf{0.987} \\
\midrule
\multicolumn{8}{c}{Protocol \textbf{ICMP}} \\
\midrule
 Random & 0.825 & 0.997 & 0.956 & 0.832 & 1.000 & 1.000 & 0 \\
  NetShare & 0.189 & 0.224 & 0.212 & - & - & - & - \\
 NetDif & 0.318 & 0.525 & 0.522 & 0.142 & 0.323 & 0.386 & 0.532 \\
 MV & 0.056 & 0.106 & 0.103 & 0.046 & 0.094 & 0.129 & 0.882 \\
 MP & 0.029 & 0.073 & 0.072 & 0.072 & 0.156 & 0.202 & 0.917 \\
 DSTF & \textbf{0.027} & \textbf{0.072} & \textbf{0.069} & \textbf{0.032 }& \textbf{0.075 }& \textbf{0.089} & \textbf{1.000} \\
\bottomrule
\end{tabular}
}
\end{table}

\noindent \textbf{Comparison with Previous Methods}: The performance of various models are shown in Table \ref{tab:MvP-sta-sim}. 
We report the results by protocol. Here, MV and MP are ablation experiments of DSTF-Diffusion, and the results are analyzed in detail in Section~\ref{sec:abla_study}. It is evident that DSTF-Diffusion achieves the best statistical similarity among all models. Moreover, as shown in the Table \ref{tab:MvP-sta-sim}, NetDiffusion only produces satisfactory results with TCP protocol data. For UDP and ICMP protocols, its statistical similarity is on a magnitude that even approaches that of random generation.
\subsection{Ablation Study}
\label{sec:abla_study}
The ablation study in Table \ref{tab:MvP-sta-sim} examines the contributions of the field stream: MV (multi-view with number mapping), MP (single-view with color mapping), and DSTF-Diffusion. Specifically, MV maps network terminologies into numerical values similar to NetDiffusion but employs a multi-view representation, while MP maps these terminologies into color space, focusing on a single view. Results confirm that both components are essential, enabling diffusion model to effectively understand network terminologies.

\begin{table}[ht]
\centering
\caption{Experimental results with different mapping methods, evaluate on protocol UDP.}
\begin{tabular}{c|c|c|c|c|c|c}
\hline
Method & \multicolumn{3}{c|}{\textbf{All Features}} & \multicolumn{3}{c}{\textbf{Protocol}} \\
 & \textbf{JSD} & \textbf{TVD} & \textbf{HD} & \textbf{JSD} & \textbf{TVD} & \textbf{HD} \\
\hline
Rand Num & 0.52 & 0.83 & 0.82 & 0.37 & 0.67 & 0.70 \\
Rand Noun & 0.10 & 0.19 & 0.17 & 0.38 & 0.44 & 0.53 \\
Rand Color & 0.08 & 0.16 & 0.19 & 0.10 & 0.38 & 0.41 \\
DSTF & \textbf{0.05} & \textbf{0.10} & \textbf{0.12 }& \textbf{0.07} & \textbf{0.20} & \textbf{0.20} \\
\hline
\end{tabular}

\label{tab:map_abla}
\end{table}

Table~\ref{tab:map_abla} presents the ablation study results for different mapping methods, measured using JSD, TSD, HD on protocol UDP. Compared to the random numeric mapping (``Rand num") used by NetDiffusion, employing semantic mappings such as random nouns (``Rand Noun") or random colors (``Rand Color") significantly reduces JSD, suggesting that semantic information contributes positively to model performance. Moreover, the proposed DSTF-Diffusion achieves the lowest JSD value of 0.05, outperforming all other mapping strategies. These results indicate that DSTF-Diffusion effectively captures underlying semantic structures within data.


\subsection{Offline DDoS Attack Classification Performance}

In this section, unless otherwise specified, we split the real attack data into training and testing sets in 8:2 ratio and augment the training set with data generated by DSTF-Diffusion in all experiments.

To validate the effectiveness of synthetic traffic for data augmentation in machine learning-based enhancements, we employ three commonly used models in the domain of traffic classification: Random Forest (RF), Decision Tree (DT), and Support Vector Machine (SVM). We test following scenarios:

\begin{itemize}
    
    \item \textbf{Data Hungry}: Here, we construct a more realistic scenario, where the task is to detect attacks (testing set) without sufficient labeled data (training data). We simulate training sets ranging from 10\% to 90\% of the size of the testing sets. For each experiment, we augment the training set with generated data until it matches the size of the testing set. This maintains the total volume of training data constant for each experiment.

    \item  \textbf{Data Imbalance}. In this scenario, we assume that the available labeled data has an unbalanced class distribution, which could potentially lead to model overfitting. Two sets of data with unbalanced class distributions (with variances of 10\% and 20\%, respectively) are created through random sampling, while the variance of testing data is 0. Subsequently, data augmentation from the generated datasets is employed to balance the data until the variance in class distribution reaches 0.
\end{itemize}

\begin{figure}[ht]
\centering
\includegraphics[width=0.40\textwidth]{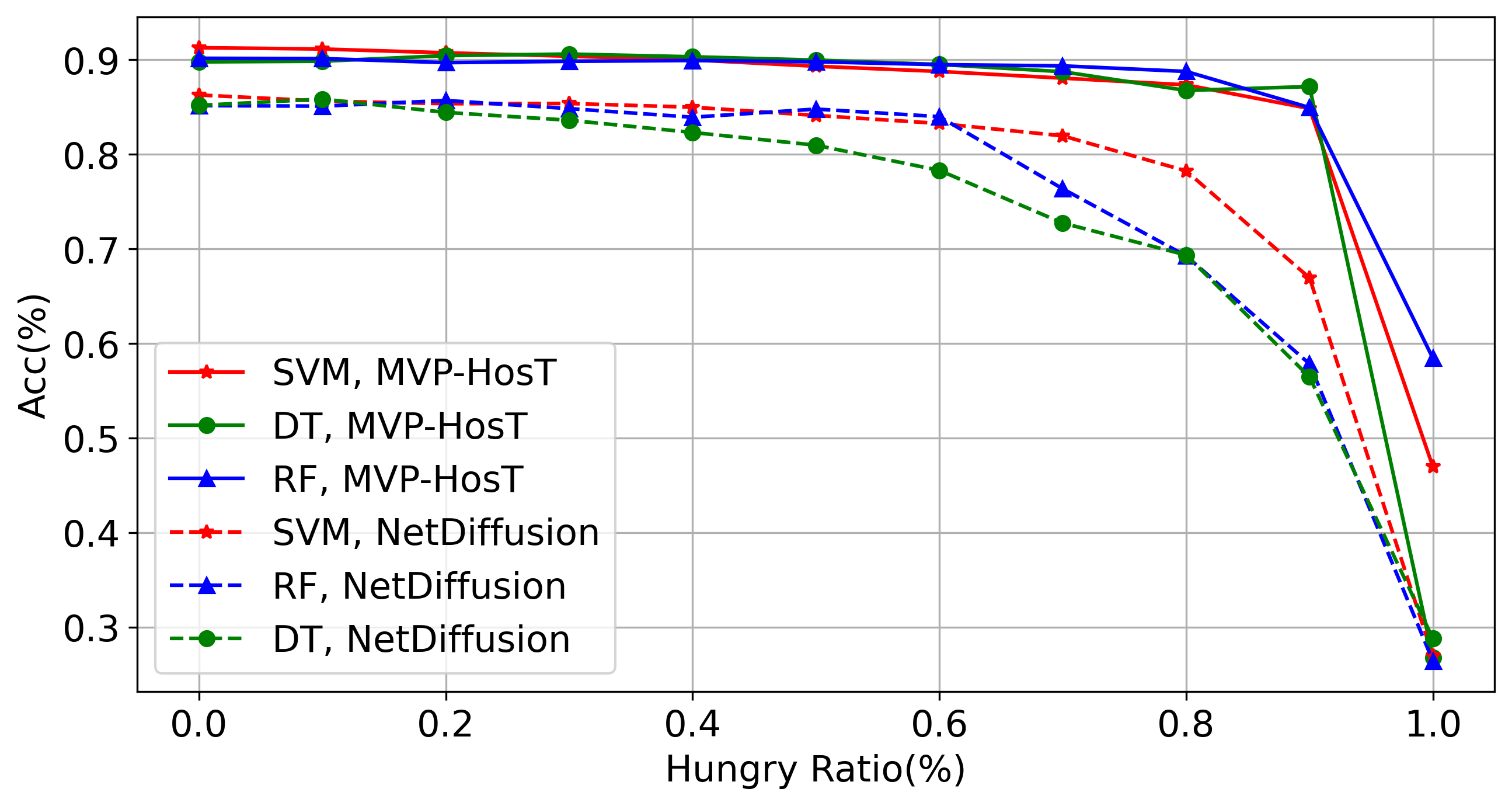} 
\caption{Results on Data Hungary.}
\label{fig:data_hungary}
\end{figure}

\noindent \textbf{Results on Data Hungry}. 
We introduce the \textit{hungry rate} to denote the ratio of training data to testing data before data augmentation, represented as $ 1 - \frac{len(\text{training\_data})}{len(\text{testing\_data})} $. When the hungry rate approaches 1, it indicates an almost complete lack of available labeled data. Figure \ref{fig:data_hungary} illustrates the change in model prediction accuracy as the hungry rate varies. It is evident that as the hungry rate increases, the model struggles to learn useful representations. However, with data augmentation, DSTF-Diffusion still enables the model to maintain an average accuracy of 85.65\% at a hungry rate of 0.9. Under the same settings, the average accuracy using NetDiffusion is  60.42\%. Furthermore, NetDiffusion only achieves a comparable accuracy to DSTF-Diffusion at a hungry rate of 0.9 when its hungry rate is 0.3. To elaborate, DSTF-Diffusion requires only 10\% of the training data to achieve the effects that NetDiffusion requires 70\% of the training data for, demonstrating that DSTF-Diffusion can improve the data hunger issue by \textbf{7 times} compared to NetDiffusion.

\noindent \textbf{Results on Data imbalance}. 
Due to the privacy concerns associated with network data, users are often reluctant to share their data, leading to imbalances when relying on a single data source, especially evident in network attacks. Imbalanced datasets can cause model overfitting. Our experiments demonstrate how DSTF-Diffusion can enhance such imbalanced datasets, as shown in the Table \ref{tab:data_imb}. Under training and testing with imbalanced data, the model's accuracy appears artificially high. However, when this model is applied to other scenarios, its performance significantly declines. The results show that models trained on augmented datasets perform more robustly, maintaining high accuracy when transferred to other scenarios. This indicates that DSTF-Diffusion can effectively train robust models even in scenarios with data imbalances, significantly reducing the consumption of training resources.

\begin{table}[ht]
\centering
\caption{Results on Data imbalance}
\label{tab:data_imb}
\resizebox{.90\columnwidth}{!}{%
\begin{tabular}{l|l|l|c|c}
\toprule
Aug. Method & Variance & Model & Before $\rightarrow$ After balance & $\Delta$ Acc \\
\midrule
NetDiffusion & 10\% & SVM & 89.62 $\rightarrow$ 89.71 & 0.09\% \\
 &  & DT & 91.64 $\rightarrow$ 92.04 & 0.40\% \\
 &  & RF & 93.43 $\rightarrow$ 95.60 & 2.17\% \\
 \midrule
DSTF & 10\% & SVM & 89.62 $\rightarrow$ 93.64 & \textbf{4.02\%} \\
 &  & DT & 91.64 $\rightarrow$ 92.43 & \textbf{0.79\%} \\
 &  & RF & 93.43 $\rightarrow$ 96.03 &  \textbf{2.60\%} \\
\midrule
NetDiffusion & 20\% & SVM & 90.97 $\rightarrow$ 90.97 & 0 \\
 &  & DT & 91.34 $\rightarrow$ 91.54 & 0.20\% \\
 &  & RF & 93.37 $\rightarrow$ 94.20 & 0.83\% \\
 \midrule
DSTF & 20\% & SVM & 90.97 $\rightarrow$ 93.71 & \textbf{2.74\%} \\
 &  & DT & 91.34 $\rightarrow$ 92.74 & \textbf{1.40\%} \\
 &  & RF & 93.37 $\rightarrow$ 95.49 &  \textbf{2.12\%} \\
\bottomrule
\end{tabular}}
\end{table}

\subsection{Real-time, Real-world Attack Detection Performance}
\label{sec:real_time}
\begin{figure}[ht]
\centering
\includegraphics[width=0.41\textwidth]{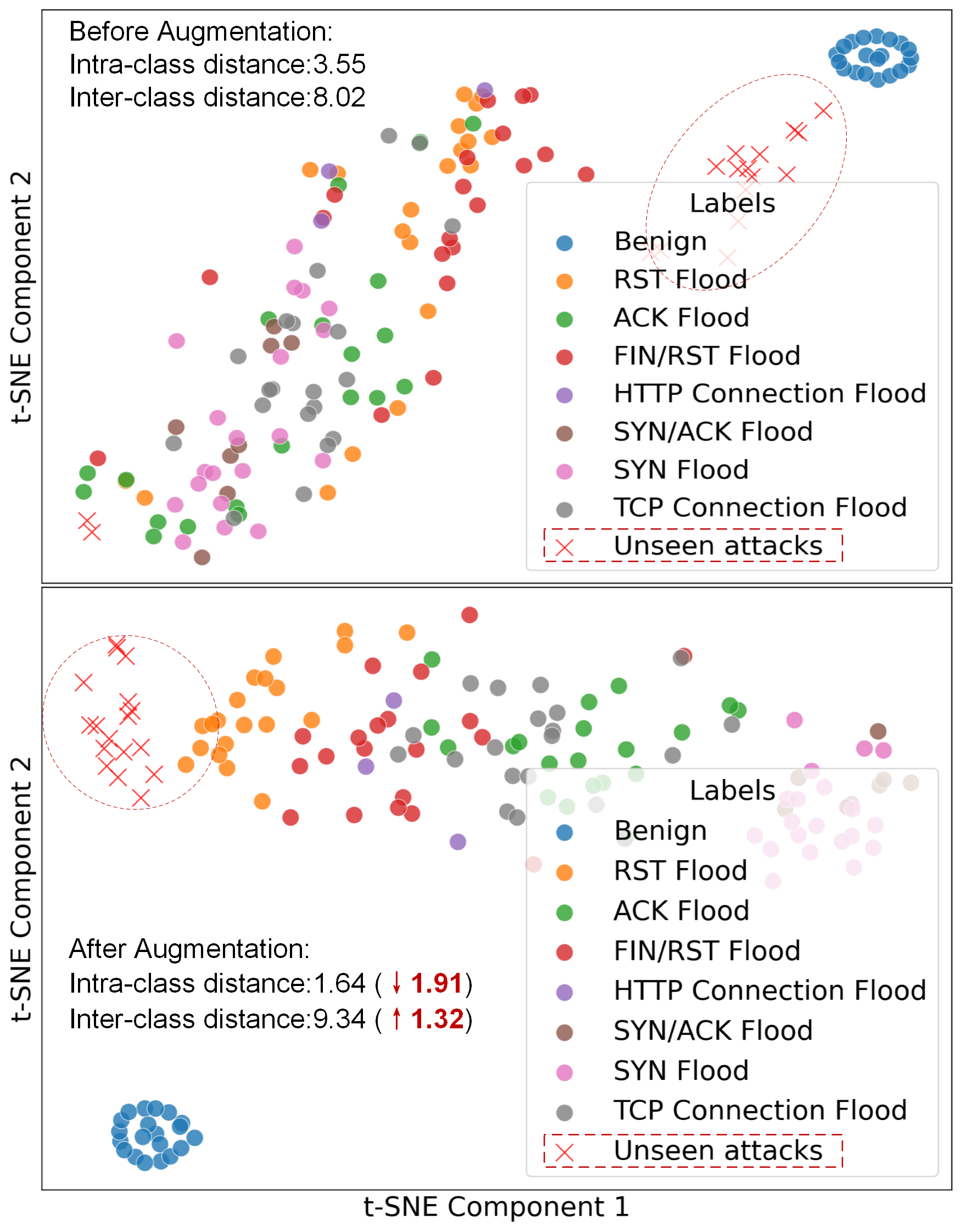} 
\caption{Feature Distribution before (Top) and after (Bottom) Augmentation.}
\label{fig:feat_dis}
\end{figure}

To validate DSTF-Diffusion in realistic network scenarios, we collected approximately 1.1 M benign and 160 K malicious traffic flows with timestamps ranging from June 20, 2022 to March 19, 2023. Attack types span eight broad categories, including RST flood, ACK flood, SYN/ACK flood, and others (see Figure~\ref{fig:feat_dis}), while temporal behavior correspond to carpet-bombing and pulse-wave attacks.
We implement a three-layer Convolutional Neural Network (CNN), which is widely applied in DDoS attack detection \cite{b3, b26}.

We tested the Combination methods designed in Section \ref{sec:comb}, including random, Markov-based, and imitative method.
As shown in Table \ref{tab:real-time}, we report two accuracy metrics: Top-1 ACC, which represents the attack type that occurs most frequently within a given period identified by the model, and overall accuracy (ACC), which represents the overall accuracy with which the model identifies all attack types within that period. It can be observed that all three sequence combination methods implemented in DSTF-Diffusion provide different benefits to the model. The random method, in particular, contributes the most significant improvement in both Top-1 ACC and ACC, reaching 97.26\% and 84.61\%, respectively. This is because the random organization method provides the most diverse data.

\begin{table}[ht]
\centering
\caption{Comparison of Real-Time Attack Detection Accuracy With Combination Methods}
\label{tab:real-time}
\begin{tabular}{c|ll}
\toprule
 Comb. Method & ACC & Top-1 ACC \\
\midrule
w/o & 77.05\% & 89.53\% \\
Imitative & 82.74\% (\textbf{+5.69\%}) & 96.30\% (\textbf{+6.77\%}) \\
Markov & 82.96\% (\textbf{+5.91\%}) & 96.84\% (\textbf{+7.31\%}) \\
Random & 84.61\% (\textbf{+7.56\%}) & 97.26\% (\textbf{+7.73\%}) \\

\bottomrule
\end{tabular}
\end{table}

In Figure \ref{fig:feat_dis}, we project features from the CNN’s output layer into two dimensions via t-SNE to compare their distributions before and after augmentation. Quantitatively, the average intra-class and inter-class distances improve from 3.55 and 8.02 to 1.64 ($\boldsymbol{\downarrow 1.91}$) and 9.34 ($\boldsymbol{\uparrow 1.32}$), respectively. These results indicate that DSTF-Diffusion-generated samples both compress intra-class variance and expand inter-class separation, thereby markedly enhancing the quality of the model’s learned representations.


\noindent \textbf{Evaluation on Unseen Attack Types.} Real-world deployment often entails previously unseen variants. As shown in Fig.~\ref{fig:feat_dis}, DSTF augmentation expands the feature-space distance between unseen attack samples and benign traffic by +4.58 (+59.25\%). This generalization arises from two complementary mechanisms: the \textbf{Temporal Stream}’s \textbf{disassembly–generation–recombination} pipeline (Sec.~\ref{sec:math:host}), which decomposes traces via STL and DTW clustering and recombines patterns to synthesize novel time series; and the \textbf{Field Stream}’s \textbf{color-based mapping} and \textbf{multi-view prompting} (Sec.~\ref{sec:math:MvP}), which produces semantically robust packet representations for unseen fields combinations. Together, these strategies drive the pronounced feature-space separation and robust performance on novel DDoS variants.

\subsection{Inference time}
As shown in Table~\ref{tab:diffusion-comparison}, using NetDiffusion’s parameter count and inference time as the baseline (1×), DSTF-Diffusion requires only 1.00017 times the parameters and 1.04 times the inference latency relative to this benchmark. This minimal overhead is accompanied by substantial improvements in the representation of complex temporal patterns and more accurate modeling of attack behaviors. The slight increase in costs is thus justified by these significant performance enhancements, clearly demonstrating the effectiveness of DSTF-Diffusion for advanced DDoS analysis tasks.
\begin{table}[htbp]
  \centering
  \caption{Comparison of DSTF-Diffusion and NetDiffusion}
  \label{tab:diffusion-comparison}
  \begin{tabular}{lcc}
    \toprule
    Method                  & Parameter Count & Inference Time \\
    \midrule
    NetDiffusion& 1.00×               & 1.00×               \\
    DSTF-Diffusion      & 1.00017×             & 1.04×               \\
    \bottomrule
  \end{tabular}
\end{table}

\section{Related Work}

The use of generative models for network data augmentation has become increasingly prominent as a method to enhance the robustness of cybersecurity systems \cite{b56}. Generative Adversarial Networks (GANs) have been pivotal in network data augmentation \cite{b43}, generating synthetic data that mimics real network behaviors for enhanced cybersecurity. While effective, the adversarial nature of GANs often limits the diversity of generated data, potentially missing novel attack patterns not present in the training set \cite{b25}.

Diffusion models \cite{b34} offer a promising alternative, known for producing high-quality and diverse samples. These models gradually transform noise into realistic data through a reverse diffusion process. Their ability to handle complex data distributions makes them suitable for generating network traffic that includes rare or emerging attack tactics \cite{b20, b37, b55}.

However, the challenge of accurately modeling the temporal dynamics of network traffic remains. Time-series models, successful in other domains \cite{b19}, have not yet provided a unified framework for network traffic time sequences. The inherent complexity of network events, especially under attack conditions, makes temporal sequence generation an open problem in network security research.

\section{Conclusion}

This work introduced DSTF-Diffusion, a novel framework for enhancing network security through packet-level DDoS data augmentation. We developed a multi-view, multi-stream network traffic generative model that employs a combination of field stream and temporal stream to address critical challenges in network data generation. Our approach not only improves data fidelity by accurately synthesizing complex network behaviors but also significantly enhances data diversity to better train and refine network security models. Through extensive experiments, our generated data demonstrated superior performance in enhancing the robustness and accuracy of network security systems against diverse and evolving attack vectors. The DSTF-Diffusion framework sets a new benchmark for research in network data augmentation.

\noindent \textbf{Limitations} DSTF-Diffusion remains bound by Stable Diffusion’s fixed input size and implicit modeling of inter-field relations, restricting synthesis to 1,024-packet windows and potentially undermining temporal continuity and semantic coherence across segments. Implicit modeling of inter-field relationships relies entirely on Stable Diffusion’s generative and semantic prowess.

\section{acknowledgment}
The work was supported by the National Key R\&D Program of China under Grant
2023YFB2904100.

\end{document}